\documentclass[notitlepage,a4paper,aps,prd,onecolumn,superscriptaddress,nofootinbib,groupedaddress,longbibliography]{revtex4-1}
\usepackage{amsmath}
\usepackage{amsfonts}
\usepackage{amssymb}
\usepackage{mathtools}
\usepackage[utf8]{inputenc}
\usepackage[T1]{fontenc}
\usepackage[colorlinks=true]{hyperref}

\newcommand{\order}[2]{\overset{\mathclap{\scriptscriptstyle #2}}{#1}\vphantom{#1}}
\newcommand{\lc}[1]{\overset{\circ}{#1}\vphantom{#1}}
\newcommand{\dd}{\mathrm{d}}

\begin{document}

\title{Post-Newtonian limit of generalized scalar-torsion theories of gravity}

\author{Kai Flathmann}
\email{kai.flathmann@uni-oldenburg.de}
\affiliation{Institut f\"ur Physik, Universit\"at Oldenburg, 26111 Oldenburg, Germany}

\author{Manuel Hohmann}
\email{manuel.hohmann@ut.ee}
\affiliation{Laboratory of Theoretical Physics, Institute of Physics, University of Tartu, W. Ostwaldi 1, 50411 Tartu, Estonia}

\begin{abstract}
In this article we derive the post-Newtonian limit of a class of teleparallel theories of gravity, where the action is a free function $L(T,X,Y,\phi)$ of the Torsion scalar $T$ and scalar quantities $X$ and $Y$ built from the dynamical scalar field $\phi$. We restrict the analysis to a massless scalar field in order to use the parameterized post-Newtonian formalism without modifications, such as introducing an effective gravitational constant which depends on the distance between the interacting masses.
In particular the results show a class of fully-conservative theories of gravity, where the only non-vanishing parameters are $\gamma$ and $\beta$. For a particular choice of the function $L(T,X,Y,\phi)$ the theory cannot be distinguished from General Relativity in its post-Newtonian approximation.
\end{abstract}

\maketitle


\section{Introduction}\label{sec:intro}
General relativity, being the most well-established and successful theory of gravity, is challenged by a number of open questions in modern physics. One of these challenges is given by cosmological observations, such as the accelerating expansion of the Universe at present and early times in its history, known as dark energy and inflation, as well as the presence of an unknown, dark matter component, which is apparent only by its gravitational effects. A potential explanation of these observations is given by modified gravity theories, of which a large and well-studied class is constituted by scalar-tensor gravity theories~\cite{Faraoni:2004pi,Fujii:2003pa}. These theories have in common that they contain one or more scalar fields, which in general is non-minimally coupled to the metric of spacetime. The gravitational dynamics of the theory is then determined through the curvature of the Levi-Civita connection of the metric, as well as the dynamics of the scalar fields.

Another issue of rather theoretical nature is our lack of understanding of the quantum behavior of gravity and its relation to the other fundamental forces present in the standard model of particle physics. While the latter are described by gauge theories, the gauge aspect is less obvious in the standard formulation of general relativity through the curvature of spacetime. However, equivalent formulations exist in which the action becomes more similar to a Yang-Mills type action, and in which curvature is replaced by either torsion or nonmetricity, or even both at the same time~\cite{BeltranJimenez:2019tjy,Jimenez:2019ghw}. Here we will focus on so-called teleparallel models of gravity, where the gravitational interaction is attributed not to the curvature of the Levi-Civita connection, but to the torsion of a flat connection~\cite{Einstein:1928,Moller:1961,Aldrovandi:2013wha,Maluf:2013gaa,Golovnev:2018red}. For the teleparallel equivalent of general relativity (TEGR) one conventionally assumes a fixed, vanishing spin connection, as it does not contribute to the field equations. However, for modified theories this approach potentially leads to the issue of local Lorentz symmetry breaking~\cite{Li:2010cg,Sotiriou:2010mv}, as spurious degrees of freedom may appear~\cite{Li:2011rn,Ong:2013qja,Izumi:2013dca,Chen:2014qtl}. This can be overcome by making use of a covariant approach including an arbitrary, flat, metric-compatible spin connection~\cite{Krssak:2015oua,Krssak:2018ywd,Bejarano:2019fii}. An alternative ansatz is the Palatini formulation~\cite{BeltranJimenez:2018vdo}.

Combining the two aforementioned approaches of scalar field extensions and teleparallel gravity, one arrives at the notion of scalar-torsion gravity theories~\cite{Hohmann:2018vle,Hohmann:2018dqh,Hohmann:2018ijr}. Various models within this class have been studied in order to address the challenges faced by general relativity~\cite{Geng:2011aj,Izumi:2013dca,Chakrabarti:2017moe,Otalora:2013tba,Jamil:2012vb,Chen:2014qsa,Bahamonde:2015hza,Bamba:2013jqa,Nojiri:2017ncd}, and a Lorentz covariant formulation has recently been put forward~\cite{Hohmann:2018rwf}. A large class of such scalar-torsion theories, for which the name $L(T, X, Y, \phi)$ theories has been introduced, is defined by a Lagrangian which is a free function of four scalar quantities derived from the torsion of the teleparallel geometry and the scalar field~\cite{Hohmann:2018dqh}. This is the class of theories we will focus on in this article.

Besides addressing the challenging issues mentioned above, a viable gravitational theory must also pass any tests on local scales, and thus in particular correctly describe the motions in our solar system. A widely used framework for such a check is the parameterized post-Newtonian (PPN) formalism~\cite{Will:1993ns,Will:2014kxa,Will:2018bme}. It characterizes gravity theories by a set of ten parameters, which have been measured with high precision in various solar system experiments.
Through the availability of this high precision data, the PPN formalism has become an important tool for assessing the viability of gravity theories.

In order to calculate the post-Newtonian limit of scalar-torsion gravity, an adaptation to theories based on a scalar field and a tetrad is required~\cite{Hayward:1981bk}. In the context of extended teleparallel gravity, such an analysis has first been performed for the original teleparallel dark energy model~\cite{Li:2013oef} and later extended to general coupling functions and potentials~\cite{Chen:2014qsa}. It turned out that the post-Newtonian limit of these theories is identical to that of general relativity. Further including a non-minimal coupling to the teleparallel boundary term, however, leads to a different post-Newtonian limit~\cite{Sadjadi:2016kwj,Emtsova:2019qsl}. In this article we make use of a recently developed extension of the PPN formalism to covariant teleparallel gravity theories~\cite{Ualikhanova:2019ygl}, and use it in order to generalize the analysis present in the aforementioned works. This allows us to derive the post-Newtonian limit of the general $L(T, X, Y, \phi)$ class of scalar-torsion theories of gravity mentioned earlier~\cite{Hohmann:2018dqh}.

The outline of this article is as follows. We start with a brief review of the dynamical variables and field equations of the class of scalar-torsion theories we consider in section~\ref{sec:dynamics}. Another brief review of the PPN formalism is presented in section~\ref{sec:ppn}, together with its adaptation to scalar-torsion gravity. We then come to the main part of the paper, which is the calculation of the post-Newtonian tetrad components leading to the derivation of the PPN parameters shown in section~\ref{sec:solution}. The resulting post-Newtonian metric and PPN parameters, which can be used for a comparison with observations and a possible restriction of the free function of the theory, are displayed in section~\ref{sec:metpar}. We apply our results to a few example theories in section~\ref{sec:examples}, before we conclude with a discussion and outlook in section~\ref{sec:conclusion}.

In this article we use uppercase Latin letters \(A, B, \ldots = 0, \ldots, 3\) for Lorentz indices, lowercase Greek letters \(\mu, \nu, \ldots = 0, \ldots, 3\) for spacetime indices and lowercase Latin letters \(i, j, \ldots = 1, \ldots, 3\) for spatial indices. In our convention the Minkowski metric \(\eta_{AB}\) and \(\eta_{\mu\nu}\) has signature \((-,+,+,+)\).

\section{Field variables and their dynamics}\label{sec:dynamics}
Before we analyze the post-Newtonian limit of the recently proposed class of scalar-torsion theories of gravity~\cite{Hohmann:2018dqh}, we review the action of the theory, the dynamical field content and the field equations. The theory is formulated in a covariant way~\cite{Hohmann:2018rwf}, where the dynamical fields are the tetrad \(\theta^A{}_{\mu}\), the flat Lorentz spin connection \(\omega^A{}_{B\mu}\) and the additional scalar field \(\phi\).  With the help of the tetrad we can define the metric as
\begin{equation}
g_{\mu\nu} = \eta_{AB}\theta^A{}_{\mu}\theta^B{}_{\nu}
\end{equation}
and the torsion tensor as
\begin{equation}
T^{\rho}{}_{\mu\nu} = e_A{}^{\rho}\left(\partial_{\mu}\theta^A{}_{\nu} - \partial_{\nu}\theta^A{}_{\mu} + \omega^A{}_{B\mu}\theta^B{}_{\nu} - \omega^A{}_{B\nu}\theta^B{}_{\mu}\right)\,.
\end{equation}
Here \(e_A{}^{\mu}\) is the inverse tetrad, which is defined in a way that \(\theta^A{}_{\mu}e_A{}^{\nu} = \delta_{\mu}^{\nu}\) and \(\theta^A{}_{\mu}e_B{}^{\mu} = \delta^A_B\). We can define the Levi-Civita connection \(\lc{\nabla}\) and the curvature tensors via the metric tensor defined above. Note that quantities with an empty circle are derived from the Levi-Civita connection.
We consider the following action
\begin{equation}\label{eqn:action}
S[\theta^A{}_{\mu}, \omega^A{}_{B\mu}, \phi, \chi^I] = S_g[\theta^A{}_{\mu}, \omega^A{}_{B\mu}, \phi] + S_m[\theta^A{}_{\mu}, \chi^I]\,,
\end{equation}
which splits into a matter part \(S_m\) and a gravitational action \(S_g\). The matter action depends on the tetrad and an arbitrary set \(\chi^I\) of matter fields. We furthermore assume, that the matter source is given by a perfect fluid. See a further discussion in section~\ref{sec:ppn}.  Another assumption we make, is, that the matter fields \(\chi^I\) do not couple directly to the teleparallel spin connection \(\omega^A{}_{B\mu}\) or the scalar field, and that the matter action is invariant under local Lorentz transformations. By taking into account all assumptions and performing integration by parts, we can write the variation of the matter action with respect to the dynamical fields as
\begin{equation}\label{eqn:matactvar}
\delta S_m[\theta^A{}_{\mu}, \chi^I] = \int_M\left[\Theta_A{}^{\mu}\delta\theta^A{}_{\mu} + \varpi_I\delta\chi^I\right]\theta\,\dd^4x\,.
\end{equation}
The energy-momentum tensor \(\Theta_{\mu\nu} = \theta^A{}_{\mu}g_{\nu\rho}\Theta_A{}^{\rho}\) is symmetric, as a consequence of the assumed Lorentz invariance, and the matter field equations are given by \(\varpi_I = 0\). Furthermore the determinant of the tetrad is \(\theta^A{}_{\mu}\) is denoted by \(\theta\).
We assume a gravitational action of the form
\begin{equation}\label{eqn:actiong}
S_g[\theta^A{}_{\mu}, \omega^A{}_{B\mu}, \phi] = \frac{1}{2\kappa^2}\int_ML(T, X, Y, \phi)\,\theta\,\dd^4x\,
\end{equation}
where
\begin{equation}\label{eqn:torsscal}
T = \frac{1}{2}T^{\rho}{}_{\mu\nu}S_{\rho}{}^{\mu\nu}
\end{equation}
is the Torsion scalar defined in terms of the superpotential
\begin{equation}\label{eqn:suppot}
S_{\rho\mu\nu} = \frac{1}{2}\left(T_{\nu\mu\rho} + T_{\rho\mu\nu} - T_{\mu\nu\rho}\right) - g_{\rho\mu}T^{\sigma}{}_{\sigma\nu} + g_{\rho\nu}T^{\sigma}{}_{\sigma\mu}\,.
\end{equation}
\begin{equation}\label{eqn:defx}
X = -\frac{1}{2}g^{\mu\nu}\phi_{,\mu}\phi_{,\nu}\,,
\end{equation}
denotes the kinetic term of the scalar field and
\begin{equation}\label{eqn:defy}
Y = g^{\mu\nu}T^{\rho}{}_{\rho\mu}\phi_{,\nu}\,.
\end{equation}
is a derivative coupling term.
By varying the total action \eqref{eqn:action} with respect to the tetrad we derive the tetrad field equation \cite{Hohmann:2018rwf}
\begin{align}\label{eqn:trfieldeqns}
E_{\mu\nu}=& -Lg_{\mu\nu}-2\lc{\nabla}_{\rho}\left(L_TS_{\nu\mu}{}^{\rho}\right)-L_T\left(T^{\rho}{}_{\rho\sigma}T^{\sigma}{}_{\mu\nu}+2T^{\rho}{}_{\rho\sigma}T_{(\mu\nu)}{}^{\sigma}-\frac{1}{2}T_{\mu\rho\sigma}T_{\nu}{}^{\rho\sigma}+T_{\mu\rho\sigma}T^{\rho\sigma}{}_\nu\right)-L_X\phi_{,\mu}\phi_{,\nu} \nonumber \\
&+\lc{\nabla}_{\nu}\left(L_Y\phi_{,\mu}\right)-\lc{\nabla}_{\sigma}\left(L_Y\phi_{,\rho}\right)g^{\rho\sigma}g_{\mu\nu}+L_Y\left(T_{(\mu\nu){}^{\rho}\phi_{,\rho}}+\frac{1}{2}T^{\rho}{}_{\mu\nu}\phi_{,\rho}+T^{\rho}{}_{\rho\mu}\phi_{,\nu}\right)-2 \kappa^2\Theta_{\mu\nu}=0\,,
\end{align}
and similarly with respect to the scalar field we derive the scalar field equation
\begin{equation}\label{eqn:scfieldeqns}
E_{\phi}=g^{\mu\nu}\lc{\nabla}_{\mu}\left(L_YT^{\rho}{}_{\rho\nu}-L_X\phi_{,\nu}\right)-L_{\phi}=0\,,
\end{equation}
where $L_{X,Y,T,\phi}$ is the partial derivative of the free function $L$ with respect to $X,Y,T$ and $\phi$, respectively. Note that we have set $\alpha\equiv 0$, in other words, there is no coupling between the scalar field and the matter field. These are the equations we will use in the remainder of this article. In order to solve them, we will perform a perturbative expansion of the dynamical fields. This will be discussed in the following section. In the following sections we will use the definitions and notations of ~\cite{Will:1993ns}; whereas~\cite{Will:2018bme} is using a slightly different treatment.

\section{Post-Newtonian approximation}\label{sec:ppn}
In this section we review the parameterized post-Newtonian (PPN) formalism~\cite{Will:1993ns,Will:2014kxa,Will:2018bme}, which is the main tool we are using in this article. The formalism we are using here, was used to analyze various scalar-torsion theories before~\cite{Hayward:1981bk,Ualikhanova:2019ygl,Emtsova:2019qsl}.
As already mentioned in the previous section, the matter field is given by a perfect fluid
\begin{equation}\label{eqn:tmunu}
\Theta^{\mu\nu} = (\rho + \rho\Pi + p)u^{\mu}u^{\nu} + pg^{\mu\nu}\,,
\end{equation}
with rest energy density \(\rho\), specific internal energy \(\Pi\), pressure \(p\) and four-velocity \(u^{\mu}\). The normalization of the four velocity \(u^{\mu}\) is given by \(u^{\mu}u^{\nu}g_{\mu\nu} = -1\). In order to use the PPN formalism, we have to assume that the velocity \(v^i = u^i/u^0\) of the source matter in a given reference frame is small in comparison to the speed of light. Then we can use a perturbative expansion of the dynamical fields in orders of the velocity \(\mathcal{O}(n) \propto |\vec{v}|^n\). Furthermore we use the Weitzenböck gauge \(\omega^A{}_{B\mu} \equiv 0\), which has been proposed in~\cite{Ualikhanova:2019ygl}. We expand the tetrad \(\theta^A{}_{\mu}\) around a flat diagonal background tetrad \(\Delta^A{}_{\mu} = \mathrm{diag}(1, 1, 1, 1)\)
\begin{equation}\label{eqn:tetradexp}
\theta^A{}_{\mu} = \Delta^A{}_{\mu} + \tau^A{}_{\mu} = \Delta^A{}_{\mu} + \order{\tau}{1}^A{}_{\mu} + \order{\tau}{2}^A{}_{\mu} + \order{\tau}{3}^A{}_{\mu} + \order{\tau}{4}^A{}_{\mu} + \mathcal{O}(5)\,.
\end{equation}
Furthermore we expand the scalar field \(\phi\)
\begin{equation}\label{eqn:scalarexp}
\phi = \Phi + \psi = \Phi + \order{\psi}{1} + \order{\psi}{2} + \order{\psi}{3} + \order{\psi}{4} + \mathcal{O}(5)\,.
\end{equation}
around its cosmological background value \(\Phi\), which will be assumed to be constant. We use overscript number to assign velocity orders to each term. For example \(\order{\psi}{n}\) is of order \(\mathcal{O}(n)\). If we assume a quasi-static gravitational field, then the changes over time are only induced by the dynamics of the source matter. Therefore time derivatives \(\partial_0\) have an additional velocity order \(\mathcal{O}(1)\). For the calculation of the first post-Newtonian approximation of the metric, we can neglect all velocity orders beyond the fourth order.
A more convenient way to write the tetrad perturbation \(\tau^A{}_{\mu}\) is to first lower the Lorentz index into a spacetime index with the help of the Minkowski metric \(\eta_{AB}\) and the background tetrad \(\Delta^A{}_{\mu}\). Then we introduce the tetrad perturbations as
\begin{equation}
\tau_{\mu\nu} = \Delta^A{}_{\mu}\eta_{AB}\tau^B{}_{\nu}\,, \quad
\order{\tau}{n}_{\mu\nu} = \Delta^A{}_{\mu}\eta_{AB}\order{\tau}{n}^B{}_{\nu}\,.
\end{equation}
It is not necessary to calculate all components of the tetrad and the scalar field up to fourth velocity order. In addition some of them simply vanish because of the Newtonian energy conservations or symmetry with respect to time reversal. The non-vanishing components of the field variables we have to calculate are~\cite{Emtsova:2019qsl}
\begin{equation}\label{eqn:ppnfields}
\order{\tau}{2}_{00}\,, \quad
\order{\tau}{2}_{ij}\,, \quad
\order{\tau}{3}_{0i}\,, \quad
\order{\tau}{3}_{i0}\,, \quad
\order{\tau}{4}_{00}\,, \quad
\order{\psi}{2}\,, \quad
\order{\psi}{4}\,.
\end{equation}
We expand all geometric quantities appearing in the field equations using the components listed above and the expansion~\eqref{eqn:tetradexp} up to the relevant velocity order. The perturbation of the metric around a flat Minkowski background is given by
\begin{equation}\label{eqn:metricexp}
\order{g}{2}_{00} = 2\order{\tau}{2}_{00}\,, \quad
\order{g}{2}_{ij} = 2\order{\tau}{2}_{(ij)}\,, \quad
\order{g}{3}_{0i} = 2\order{\tau}{3}_{(i0)}\,, \quad
\order{g}{4}_{00} = -(\order{\tau}{2}_{00})^2 + 2\order{\tau}{4}_{00}\,.
\end{equation}
The remaining terms calculated by using the perturbed tetrad can be found in~\cite{Ualikhanova:2019ygl}.
Next we can apply the post-Newtonian expansion to the gravitational field equations~\eqref{eqn:trfieldeqns} and~\eqref{eqn:scfieldeqns}. In order to apply it to the geometry side of the equations, we have to expand the free function $L\left(T,X,Y,\phi\right)$ and its derivatives as a Taylor series
\begin{align}\label{eqn:taylorseries}
L &= l_0+l_\phi\psi+\frac{1}{2}l_{\phi\phi}\psi^2+l_T T+l_XX+l_YY \,, \nonumber \\
L_T &= l_T+l_{T\phi}\psi+\frac{1}{2}l_{T\phi\phi}\psi^2+l_{TX}X+l_{TY}Y+l_{TT}T \,, \nonumber \\
L_X &= l_X+l_{X\phi}\psi+\frac{1}{2}l_{X\phi\phi}\psi^2+l_{TX}T+l_{XY}Y+l_{XX}X \,, \nonumber \\
L_Y &= l_Y+l_{Y\phi}\psi+\frac{1}{2}l_{Y\phi\phi}\psi^2+l_{TY}T+l_{XY}X+l_{YY}Y \,, \nonumber \\
L_\phi &= l_\phi +l_{\phi\phi}\psi+l_{T\phi}T+l_{Y\phi}Y+l_{X\phi}X+\frac{1}{2}l_{\phi\phi\phi}\psi^2 \,,
\end{align}
where the Taylor coefficients $l_0,l_{\phi},\ldots$ are calculated at the cosmological background which implies $T=X=Y=0$ and $\phi=\Phi$ and are assumed to be of velocity order \(\mathcal{O}(0)\).

Finally, for the matter side of the field equations, we must also expand the energy-momentum tensor~\eqref{eqn:tmunu} into velocity orders. For this purpose, we use the standard assignment of velocity orders also to the rest mass density, specific internal energy and pressure of the perfect fluid; based on their orders of magnitude in the solar system one assigns velocity orders \(\mathcal{O}(2)\) to \(\rho\) and \(\Pi\) and \(\mathcal{O}(4)\) to \(p\)~\cite{Will:1993ns}. The energy-momentum tensor~\eqref{eqn:tmunu} can then be expanded in the form
\begin{subequations}\label{eqn:energymomentum}
\begin{align}
\Theta_{00} &= \rho\left(1 + \Pi + v^2 - 2\order{\tau}{2}_{00}\right) + \mathcal{O}(6)\,,\\
\Theta_{0j} &= -\rho v_j + \mathcal{O}(5)\,,\\
\Theta_{ij} &= \rho v_iv_j + p\delta_{ij} + \mathcal{O}(6)\,.
\end{align}
\end{subequations}
These are all formulas which will be necessary for the post-Newtonian expansion of the field equations. We will proceed with this expansion and their solution in the following section.

\section{Solving the field equations}\label{sec:solution}
This section is devoted to finding the perturbative solution of the field equations~\eqref{eqn:trfieldeqns} and~\eqref{eqn:scfieldeqns} in the standard post-Newtonian gauge, by making use of the general formalism discussed in the preceding section. Our derivation proceeds order by order in the post-Newtonian expansion. The zeroth velocity order, which corresponds to the background solution of the vacuum field equations, is discussed in section~\ref{ssec:order0}. We then solve the field equations at the second order in section~\ref{ssec:order2}, at the third order in section~\ref{ssec:order3} and at the fourth order in section~\ref{ssec:order4}.

\subsection{Zeroth velocity order}\label{ssec:order0}
We start our derivation with the observation that at the zeroth velocity order the energy-momentum tensor vanishes, \(\order{\Theta}{0}_{\mu\nu} = 0\), so that we are left with solving the vacuum field equations. Inserting our assumed background values \(\order{\theta}{0}^A{}_{\mu} = \Delta^A{}_{\mu}\) for the tetrad and \(\order{\phi}{0} = \Phi\) into the respective field equations~\eqref{eqn:trfieldeqns} and~\eqref{eqn:scfieldeqns}, we find that their zeroth order is given by
\begin{equation}
\order{E}{0}_{00} = -l_0\,, \quad
\order{E}{0}_{ij} = l_0\delta_{ij}\,, \quad
\order{E}{0} = 4l_0l_Y-4l_Tl_\phi\,.
\end{equation}
It thus follows that the perturbation ansatz is consistent with the vacuum field equations only if the parameter functions satisfy \(l_0 = 0\). Furthermore we choose $l_{\phi}=0$ to satisfy the zeroth order scalar field equation. In the following we will restrict ourselves to theories satisfying these conditions. These conditions are, in fact, less restrictive than they might seem at first sight. The condition \(l = 0\) can be interpreted as the vanishing of the cosmological constant, or at least that its effect is negligible in the solar system. Further, \(l_{\phi} = 0\) means that the background value of the scalar field should mark an extremal value of its potential; this case can be found as an attractor in scalar-torsion cosmology, and should therefore be a valid assumption in the late universe~\cite{Jarv:2015odu}.

\subsection{Second velocity order}\label{ssec:order2}
By summing up the field equations in the following way
\begin{equation}
4l_TE_\phi-l_Yg^{\mu\nu}E_{\mu\nu}=0\,,
\end{equation}
we can decouple the second velocity order of the scalar field from the tetrad
\begin{equation}\label{eqn:scfieldeqn_order2}
\left(3l_Y^2-4l_Tl_X\right)\triangle\order{\psi}{2}= 2\kappa^2 l_Y\rho \,.
\end{equation}
Note that we restrict ourselves to $l_{\phi\phi}=0$ in order to have a massless scalar field. Equation \eqref{eqn:scfieldeqn_order2} can be solved in terms of the Newtonian potential $U$ defined by
\begin{equation}\label{eqn:PPNpotential_U}
\triangle U=-4\pi\rho\,,
\end{equation}
where $\triangle=\eta^{ab}\partial_a\partial_b$ is the flat space Laplace operator. The solution is then given by
\begin{equation}\label{eqn:sol_psi2}
\order{\psi}{2}=\frac{4l_Y}{4l_Tl_X-3l_Y^2}\frac{\kappa^2}{8\pi}U \,.
\end{equation}
For convenience we will now use the trace reversed version of the field equations \eqref{eqn:trfieldeqns}
\begin{equation}\label{eqn:trfieldeqn_tracereversed}
\bar{E}_{\mu\nu}=E_{\mu\nu}-\frac{1}{2}g_{\mu\nu}g^{\alpha\beta}E_{\alpha\beta}\,.
\end{equation}
This will be beneficial in particular for solving the fourth order tetrad field equations. The relevant components are
\begin{align}\label{eqn:trfieldeqn_order2}
\bar{E}_{00} &=2l_T\triangle\order{\tau}{2}_{00}-\kappa^2\rho-\frac{1}{2}l_Y\triangle\order{\psi}{2} \nonumber\\
\bar{E}_{ij} &=2l_T\triangle\order{\tau}{2}_{(ij)}-\kappa^2\rho\delta_{ij}+\frac{1}{2}l_Y\delta_{ij}\triangle\order{\psi}{2}+l_Y\order{\psi}{2}_{,ij}-2l_T\order{\tau}{2}_{00,ij}+2l_T\order{\tau}{2}_{kk,ij}-2l_T\order{\tau}{2}_{k(i,kj)}-2l_T\order{\tau}{2}_{(ik,kj)}\,.
\end{align}
By substituting the solution for the second order scalar field $\order{\psi}{2}$ (Eq. \eqref{eqn:sol_psi2}) in equation \eqref{eqn:trfieldeqn_order2} we can solve for the tetrad components in terms of the Newtonian potential defined in equation \eqref{eqn:PPNpotential_U}
\begin{align}\label{eqn:sol_tau2}
\order{\tau}{2}_{00} &= \frac{4\left(l_Y^2-l_Tl_X\right)}{l_T\left(4l_Tl_X-3l_Y^2\right)}\frac{\kappa^2}{8\pi}U\nonumber\\
\order{\tau}{2}_{ij} &=\frac{2\left(l_Y^2-2l_Tl_X\right)}{l_T\left(4l_Tl_X-3l_Y^2\right)}\frac{\kappa^2}{8\pi}U\delta_{ij}\,.
\end{align}

\subsection{Third velocity order}\label{ssec:order3}
In the third velocity order the only non-vanishing components of the field equations are $\bar{E}_{i0}$ and $\bar{E}_{0i}$. To solve these equations we introduce the vector potentials $V_i$ and $W_i$ defined by
\begin{equation}\label{eqn:PPNpotentials_ViandWi}
\triangle V_i=-4\pi\rho v_i\,, \qquad \triangle W_i=-4\pi\rho v_i+2 U_{,0i}\,.
\end{equation}
It is also convenient to use the combination
\begin{align}\label{eqn:trfieldeqn_order3}
\bar{E}_{i0}+\bar{E}_{0i}=\triangle\order{\tau}{3}_{(i0)}+4\kappa^2\rho v_i+2l_Y\order{\psi}{2}_{,0i}-l_T\order{\tau}{2}_{(ki),0k}+4l_T\order{\tau}{2}_{kk,0i}-l_T\order{\tau}{3}_{(0k),ki}\,,
\end{align}
which can be solved by introducing a free parameter $a_0$
\begin{equation}\label{eqn:sol_tau3}
\order{\tau}{3}_{i0}=\order{\tau}{3}_{0i}=\frac{\kappa^2}{8\pi}\left(a_0V_i+\left(\frac{2}{l_T}-a_0\right)W_i\right)\,.
\end{equation}
The constant parameter $a_0$ will be determined by demanding the standard PPN gauge for the solution of the fourth order tetrad $\order{\tau}{4}_{00}$.

\subsection{Fourth velocity order}\label{ssec:order4}
The remaining fourth order of the $(00)$-component of the tetrad can be found by summing up the field equations in the following way
\begin{equation}\label{eqn:trfieldeqn_order4}
\left(3l_Y^2-4l_Tl_X\right)\bar{E}_{00}+\frac{1}{2}l_Y\left(4l_TE_{\phi}-l_Yg^{\mu\nu}E_{\mu\nu}\right)=0 \,.
\end{equation}
To solve this equation we make an ansatz for the tetrad
\begin{equation}\label{eqn:sol_tau4a}
\order{\tau}{4}_{00}=\frac{\kappa^2}{8\pi}\left(a_1\Phi_1+\frac{\kappa^2}{8\pi}a_2\Phi_2+a_3\Phi_3+a_4\Phi_4+\frac{\kappa^2}{8\pi}a_4U^2\right)\,,
\end{equation}
where the $\Phi_i$ are the typical PPN potentials defined by
\begin{equation}
\triangle\Phi_1= -4\pi\rho v^2\,, \qquad
\triangle\Phi_2= -4\pi\rho U\,, \qquad
\triangle\Phi_3= -4\pi\rho \Pi\,, \qquad
\triangle\Phi_4= -4\pi p\,.
\end{equation}
Inserting $\order{\tau}{4}_{00}$ and all lower order tetrad components into Eq. \eqref{eqn:trfieldeqn_order4} leads to a system of algebraic equations for $a_0$ to $a_5$ with the solution
\begin{subequations}
\begin{align}\label{eqn:sol_tau4b}
a_0 &= \frac{7l_Tl_X-5l_Y^2}{l_T\left(4l_Tl_X-3l_Y^2\right)}\,,\\
a_1 &= -\frac{2}{l_T} \,,\\
a_2 &= \frac{4\left(32l_T^3l_X^3+8l_T^2l_{T\phi}l_X^2l_Y^2-68l_T^2l_{T\phi}l_Y^2-16l_Tl_{T\phi}l_Xl_Y^2-2l_T^2l_{X\phi}l_Y^3+45l_Tl_Xl_Y^4+6l_{T\phi}l_Y^5-9l_Y^6+4l_T^2l_Xl_Y^2l_{Y\phi}\right)}{l_T^2\left(4l_Tl_X-3l_Y^2\right)} \,,\\
a_3 &= \frac{4\left(l_Y^2-l_Tl_X\right)}{l_T\left(4l_Tl_X-3l_Y^2\right)} \,, \\
a_4 &= \frac{6\left(l_Y^2-2l_Tl_X\right)}{l_T\left(4l_Tl_X-3l_Y^2\right)} \,, \\
a_5 &= -\frac{2\left(16l_T^3l_X^3-8l_T^2l_{T\phi}l_X^2l_Y^2-40l_T^2l_{T\phi}l_Y^2+16l_Tl_{T\phi}l_Xl_Y^2+2l_T^2l_{X\phi}l_Y^3+33l_Tl_Xl_Y^4-6l_{T\phi}l_Y^5-9l_Y^6-4l_T^2l_Xl_Y^2l_{Y\phi}\right)}{l_T^2\left(4l_Tl_X-3l_Y^2\right)}\,.
\end{align}
\end{subequations}

\section{PPN metric and Parameters}\label{sec:metpar}
With the tetrad components calculated in the previous section, we have now solved the field equations to the necessary order so that we can construct the post-Newtonian metric, which we show in section~\ref{ssec:metric}. From this we read off the PPN parameters in section~\ref{ssec:param}.

\subsection{PPN metric}\label{ssec:metric}
We can now substitute the determined tetrad components (Equations \eqref{eqn:sol_tau2}, \eqref{eqn:sol_tau3} and \eqref{eqn:sol_tau4a} with \eqref{eqn:sol_tau4b}). By defining the gravitational constant as $G=\frac{\kappa^2}{2\pi}\frac{l_Y^2-l_Tl_X}{l_T\left(4l_Tl_X-3l_Y^2\right)}$ and then setting $G=1$, we derive the metric components in the standard PPN form
\begin{subequations}
\begin{align}
\order{g}{2}_{00} =\ &2U \,, \\
\order{g}{2}_{ij} =\ &\frac{2l_Tl_X-l_Y^2}{l_Tl_X-l_Y^2}U\delta_{ij} \,,\\
\order{g}{3}_{i0} =\ &\frac{5l_Y^2-7l_Tl_X}{2\left(l_Tl_X-l_Y^2\right)}V_i-\frac{1}{2}W_i \,,\\
\order{g}{4}_{00} =\ &\left\{\left[32l_T^3l_X^3-3l_Y^5\left(2l_{T\phi}+7l_Y\right)+l_Tl_Xl_Y^3\left(16l_{T\phi}+73l_Y\right)-8l_T^2l_Yl_{T\phi}l_X^2-2l_T^2l_Y^3\left(42l_X^2-l_{X\phi}l_Y+2l_Xl_{Y\phi}\right)\right]U^2 \right.\nonumber\\
&+\left[-16l_T^3l_X^3l_Tl_Xl_Y^3\left(8l_{T\phi}-\frac{45}{2}l_Y\right)-3l_{T\phi}l_Y^5+\frac{9}{2}l_Y^6+l_T^2l_Y\left(-4l_{T\phi}l_X^2+34l_Yl_X^2+l_Y^2l_{X\phi}-2l_Xl_Yl_{Y\phi}\right)\right]\Phi_2\nonumber\\
&+\left.\left(3l_Y^2-4l_Tl_X\right)\Phi_1+2\Phi_3+3\left(l_Y^2-2l_Tl_X\right)\Phi_4\right\}\frac{1}{\left(4l_Tl_X-3l_Y^2\right)\left(l_Y^2-l_Tl_X\right)^2} \,.
\end{align}
\end{subequations}
Our definition of the gravitational constant $G$ can be compared with the effective gravitational constant $G_{\text{eff}}$ in Equation (31) in reference \cite{Jarv:2017npl} for a massless scalar field ($m_{\Psi}=0$), if we choose $l_Y=1$, $l_X=\omega/2\Psi_0$ and $l_T=-\Psi_0$, which are the values one would obtain by rewriting the action given in Equation (2) in the same reference \cite{Jarv:2017npl} in the teleparallel language.

\subsection{PPN parameters}\label{ssec:param}
The PPN parameters can now be read off by comparing the metric components with the standard PPN metric~\cite{Will:1993ns,Will:2014kxa,Will:2018bme}, thus $\xi=\alpha_1=\alpha_2=\alpha_3=\zeta_1=\zeta_2=\zeta_3=\zeta_4=0$ and
\begin{align}\label{eqn:betagamma}
\gamma-1 &= \frac{l_Y^2}{2l_Tl_X-2l_Y^2}\,, \\
\beta-1 &= \frac{l_Y\left[l_Tl_Xl_Y^2\left(16l_{T\phi}-7l_Y\right)+3l_Y^4\left(l_Y-2l_{T\phi}\right)-8l_T^2l_X^2l_{T\phi}+2l_T^2l_Y\left(2l_X^2+l_Yl_{X\phi}-2l_Xl_{Y\phi}\right)\right]}{8\left(4l_Tl_X-3l_Y^2\right)\left(l_Y^2-l_Tl_X\right)^2}\,.
\end{align}
Since all of the parameters except \(\beta\) and \(\gamma\) vanish the theory is fully conservative. This means that the total energy-momentum is conserved and preferred frame or preferred location effects can not appear in this theory. We also find a class of theories which is indistinguishable from general relativity by their PPN parameters, which take the values $\gamma=\beta=1$; this is satisfied for all theories with $l_Y=0$.

To understand this result, it is useful to consult the field equations~\eqref{eqn:trfieldeqns}. These show that for \(l_Y = 0\) the scalar field becomes minimally coupled to gravity, up to the order which determines the post-Newtonian limit we calculated here. As a consequence, we see that the scalar field vanishes at the second velocity order, see equation~\eqref{eqn:sol_psi2}, as it is not sourced by the matter energy-momentum. This means that although the scalar field is present in the theory and, following our assumption of a massless field, has a long range without any exponential (Yukawa-type) suppression, it does not mediate any gravitational interaction, and therefore does not lead to any modification of the PPN parameters. However, expanding the field equations~\eqref{eqn:trfieldeqns} into higher perturbation orders, which are not relevant for the PPN limit, but may become relevant, e.g., for gravitational radiation, one may expect higher order correction terms to appear, which depend on a higher order Taylor expansion of the Lagrangian function \(L\). If such terms are present, they may be testable by gravitational experiments. However, the calculation of such higher order corrections exceeds the scope of this article, and we leave it for future work.

\section{Example theories}\label{sec:examples}
With the general result~\eqref{eqn:betagamma} at hand, we may now apply our findings to more specific theories within the general class we discussed in this article. In this section we discuss two such examples. A class of theories which is constructed similarly to scalar-curvature theories~\cite{Hohmann:2018ijr} is discussed in section~\ref{ssec:stg}. Another class of theories without derivative couplings~\cite{Hohmann:2018rwf} is discussed in section~\ref{ssec:noder}.

\subsection{Scalar-torsion analogue of scalar-curvature gravity}\label{ssec:stg}
As the first example class of theories we consider an analogue of scalar-curvature gravity theories~\cite{Hohmann:2018ijr}. The Lagrangian takes the form
\begin{equation}\label{eqn:classaction}
L(T, X, Y, \phi) = -\mathcal{A}(\phi)T + 2\mathcal{B}(\phi)X + 2\mathcal{C}(\phi)Y - 2\kappa^2\mathcal{V}(\phi)\,,
\end{equation}
and depends on four free functions \(\mathcal{A}, \mathcal{B}, \mathcal{C}, \mathcal{V}\) of the scalar field. In order to satisfy the background conditions \(l_0 = l_{\phi} = 0\) we restrict ourselves to theories satisfying \(V = V' = 0\), where we used the abbreviations \(V = \mathcal{V}(\Phi)\), \(V' = \mathcal{V}'(\Phi)\) for the Taylor coefficients. Further, to avoid any mass terms and satisfy the conditions \(l_{\phi\phi} = l_{\phi\phi\phi} = 0\), we choose \(V'' = V''' = 0\). The remaining relevant Taylor coefficients of the Lagrangian are given by
\begin{equation}
l_T = -A\,, \quad
l_X = 2B\,, \quad
l_Y = 2C\,, \quad
l_{T\phi} = -A'\,, \quad
l_{X\phi} = 2B'\,, \quad
l_{Y\phi} = 2C'\,.
\end{equation}
Inserting these values into the expression~\eqref{eqn:betagamma} for the PPN parameters then yields their values
\begin{equation}
\gamma = 1 - \frac{C^2}{AB + 2C^2}\,, \quad
\beta = 1 - \frac{C\{6C^4(C + A') + ABC^2(7C + 8A') + A^2[2B^2(C + A') + B'C^2 - 2BCC']\}}{4(AB + 2C^2)^2(2AB + 3C^2)}\,.
\end{equation}
It follows that in the case \(C = 0\) they reduce to the general relativity values \(\beta = \gamma = 1\). We also find that the result agrees with the massless case of a previous calculation of the PPN parameters for this particular class of theories~\cite{Emtsova:2019qsl}.

\subsection{Scalar-torsion theory without derivative couplings}\label{ssec:noder}
The second class of example theories we consider is based on an action which does not contain the derivative coupling term \(Y\)~\cite{Hohmann:2018rwf}. In this case the Lagrangian is given by
\begin{equation}\label{eqn:ndactiong}
L(T, X, Y, \phi) = F(T,\phi) - 2Z(\phi)X\,,
\end{equation}
with free functions \(F\) and \(Z\). Due to the absence of the derivative coupling, one immediately finds the Taylor coefficient \(l_Y = 0\). A comparison with the result~\eqref{eqn:betagamma} for the PPN parameters therefore suggests that \(\beta = \gamma = 1\), so that these theories have PPN parameters identical to that of general relativity, provided that the denominators in~\eqref{eqn:betagamma} are non-vanishing. However, it is natural to consider theories with \(l_T = F_T(0, \Phi) \neq 0\) and \(l_X = -2Z(\Phi) \neq 0\), so that in the weak field limit both the tetrad and the scalar field have non-degenerate kinetic terms and strong coupling issues are avoided. Theories satisfying these conditions indeed have PPN parameters \(\beta = \gamma = 1\).

\section{Conclusion}\label{sec:conclusion}
We have derived the post-Newtonian limit and PPN parameters for a general class of scalar-torsion theories of gravity with a massless scalar field. We represented the free function as a Taylor series around the cosmological value of the scalar field. We restrict the analysis to a massless scalar field in order to avoid Yukawa-type potentials. To guarantee an asymptotically flat Minkowski background we had to set the zeroth order coefficient to zero. Solving the field equations order by order, we derived the metric of a perfect fluid up to the first post-Newtonian order and determined the PPN parameters. All parameters other than the usual Eddington-Robertson-Schiff parameters $\gamma$ and $\beta$ are equal to zero. Therefore this class of theory predicts neither preferred frame or location effects nor the Nordvedt effect. Furthermore the total energy-momentum is globally conserved. These classes of theories are called fully-conservative. We also pointed out, that if the scalar field is minimally coupled (i.e., there is no derivative coupling, $l_Y=0$), it does not contribute to the gravitational interaction at the post-Newtonian level, and so these theories are indistinguishable from general relativity at the level of the PPN parameters.
As examples we calculated the parameters $\gamma$ and $\beta$ for the scalar-torsion analogue of scalar-curvature gravity and a scalar-torsion theory without derivative coupling in section~\ref{sec:examples}.

This work could be extended by the calculation of the parameters $\gamma$ and $\beta$ for theories with a massive scalar field, along the lines of previous works on a more specific class of scalar-torsion theories~\cite{Emtsova:2019qsl} or scalar-curvature theories~\cite{Hohmann:2013rba,Scharer:2014kya,Hohmann:2017qje}. The parameters and the gravitational constant will then depend on the spatial coordinates~\cite{Chen:2014qsa}. One may also consider scalar-torsion theories with more general couplings between the scalar field and the teleparallel geometry, such as the recently proposed teleparallel extension to Horndeski gravity~\cite{Bahamonde:2019shr} or theories obtained from disformal transformations~\cite{Hohmann:2019gmt}, thus extending previous results on the curvature formulation of Horndeski gravity~\cite{Hohmann:2015kra}. Furthermore it is interesting to analyze theories coupled to more than one massive or massless scalar field. For example the multiscalar extension of the previously analyzed theory~\cite{Hohmann:2018dqh}, following a similar treatment as in multiscalar-curvature theory~\cite{Hohmann:2016yfd}. In a similar fashion, also theories featuring nonmetricity instead of torsion coupled to scalar fields~\cite{Jarv:2018bgs,Runkla:2018xrv} may be considered.

Another interesting possibility is the calculation of the second or even higher post-Newtonian order  for this general class of theories. This could lead to a study of the emitted gravitational waves of compact objects, especially inspiralling compact binaries~\cite{Blanchet:2013haa}. In particular it may be the case that theories satisfying $l_Y=0$ are distinguishable from general relativity in a higher post-Newtonian order, if the scalar field contributes to the gravitational interaction via higher order coupling terms.

\begin{acknowledgments}
KF gratefully acknowledges support by the DFG within the Research Training Group \textit{Models of Gravity} and mobility funding from the European Regional Development Fund through \textit{Dora Pluss}. MH gratefully acknowledges the full financial support by the Estonian Research Council through the Personal Research Funding project PRG356 and by the European Regional Development Fund through the Center of Excellence TK133 ``The Dark Side of the Universe''. This article is based upon work from COST Action CANTATA (CA 15117), supported by COST (European Cooperation in Science and Technology).
\end{acknowledgments}

\bibliography{../teleppn}
\end{document}